\documentclass[preprint,preprintnumbers,amsmath,amssymb,floatfix,nofootinbib]{revtex4}
\usepackage{rotating}

\usepackage{multirow}

\usepackage[utf8]{inputenc}
\usepackage{amssymb,esvect,amsmath,graphicx,latexsym,amsthm,slashed,eso-pic,verbatim}
\usepackage{epsf, array, color}
\usepackage{graphicx}
\usepackage{subfigure}
\usepackage{bbold}
\usepackage{amsmath}
\usepackage{mathtools}
\usepackage{dcolumn}\usepackage{hyperref}
\usepackage{slashed}
\usepackage{enumerate}
\usepackage{latexsym}
\usepackage{cancel}
\usepackage[normalem]{ulem} 
\usepackage[utf8]{inputenc}

\def\ch{{\mathcal H}}
\def\cp{{\mathcal P}}
\def\cR{{\mathcal R}}

\def\cb{{\mathcal B}}
\def\cs{{\mathcal S}}
\def\cu{{\mathcal U}}
\newcommand{\ptv}{p_T^{\mathrm{veto}}}
\newcommand{\der}[1]{\mathrm{d} #1 \,}
\newcommand{\bea}{\begin{eqnarray}}
\newcommand{\eea}{\end{eqnarray}}

\def\beq{\begin{equation}}
\def\eeq{\end{equation}}

\newcommand{\co}{\mathcal{O}}

\begin{document}

\preprint {FERMILAB-PUB-16-210-PPD-T}
\preprint {YITP-SB-16-24}

\title{Resummation of Jet Veto Logarithms at partial N$^3$LL + NNLO for $W^+W^-$ Production at the LHC}

\author{S.~Dawson}
\affiliation{Department of Physics, Brookhaven National Laboratory, Upton, N.Y., 11973,  U.S.A.}
\author{P.~Jaiswal}
\affiliation{Department of Physics, Brown University, Providence, RI 02912,U.S.A}
\author{Ye Li}
\affiliation{Fermilab, PO Box 500, Batavia, IL 60510, USA}
\author{Harikrishnan Ramani}
\affiliation{C. N. Yang Institute for Theoretical Physics, Stony Brook University, Stony Brook, NY 11794, U.S.A.}
\author{Mao Zeng}
\affiliation{Department of Physics and Astronomy, University of California, Los Angeles, CA 90095, U.S.A.}

\date{\today}

\newcommand{\ww}{
$W^+W^-$
}
\begin{abstract}
We compute the resummed on-shell $W^+ W^-$ production cross section under a jet-veto at the LHC to partial  N$^3$LL order matched to the fixed order NNLO result. Differential NNLO cross sections are obtained from an implementation of $q_T$ subtraction in Sherpa. The two-loop virtual corrections to the $q \bar q \rightarrow W^+ W^-$ amplitude, used in both fixed order and resummation predictions, are extracted from the public code {\tt qqvvamp}. We perform resummation using soft collinear effective theory (SCET), with approximate beam functions where only the logarithmic terms are included at two-loop. In addition to scale uncertainties from the hard matching scale and the factorization scale, rapidity scale variations are obtained within the analytic regulator approach. Our resummation results show a decrease in the jet-veto cross-section compared to NNLO fixed order predictions, with reduced scale uncertainties compared to NNLL+NLO resummed predictions.  We include the loop-induced $gg$ contribution with jet veto resummation to NLL+LO.  The prediction shows good agreement with recent LHC measurements.
\end{abstract}

\maketitle

\section{Introduction}
\makeatletter{}Run 1 and Run 2 of the LHC represent an unprecedented reach in the energy frontier. This resulted in the Higgs discovery and also opened up opportunities for electroweak precision measurements. While precision electroweak processes are important in their own right, they also serve as the largest backgrounds to the important Higgs decays. Thus it is imperative to confirm electroweak predictions.

The WW channel is particularly interesting  because it tests the structure of the three gauge boson vertex and is extremely sensitive
to anomalous gauge boson couplings.  Furthermore, the \ww channel
is a significant  background for many beyond the standard model searches (BSM).  The $h\rightarrow W^+ W^- $ signal has the
standard model (SM)  \ww production as an irreducible background.  All of these considerations make it crucial  to understand electroweak \ww production as
accurately as possible\cite{Gehrmann:2014fva,Grazzini:2016ctr,Catani:2009sm,Frixione:1993yp,Campbell:2011bn,Dawson:2013lya}.

The 7 and 8 TeV run, while confirming the Higgs prediction, provided  little evidence in terms of BSM physics, with most electroweak channels  agreeing exactly with the SM predictions. Comparing with the NLO fixed order calculation, a 3$\sigma$ excess in the \ww
channel was initially reported by both ATLAS and CMS \cite{ATLAS:2012mec,ATLAS:2014xea,CMS:2012cva,CMS:2012zva} and this led to speculation about the existence of new physics \cite{Curtin:2013gta,Curtin:2012nn,Jaiswal:2013xra,Rolbiecki:2013fia,Kim:2014eva,Curtin:2014zua}.  Recent
experimental results at $8$\cite{Khachatryan:2015sga,Aad:2016wpd} and $13$ TeV\cite{ATLAS:2016rin,CMS:2016vww}, however, show good agreement with NNLO predictions\cite{Gehrmann:2014fva,Grazzini:2016ctr}.

It is interesting to note that the \ww channel has a large top quark background. In order to tame this background,  events containing jets with transverse momentum above $25 ~(30)$ GeV for ATLAS (CMS) are vetoed. This jet veto cut necessary to obtain the fiducial cross section is unique to the \ww channel among electroweak processes. The presence of a new scale (the jet-veto scale) different from the typical energy scale $\sim 2m_W$ introduces large logarithms in perturbation theory. In fact, the average $m_{WW}$ is about 50\% larger than the absolute threshold  $2m_W$, making the scale disparity even larger.
These effects can be significant and increase the theoretical prediction for the fiducial cross section compared with NLO parton shower predictions\cite{Nason:2013ydw,Meade:2014fca,Jaiswal:2014yba,Monni:2014zra,Becher:2014aya,Jaiswal:2015vda}. Including jet veto effects, along with the NNLO fixed order calculation of Ref.  \cite{Gehrmann:2014fva} creates better agreement with experimental results and indeed this is confirmed by CMS \cite{Khachatryan:2015sga}. These developments are summarized in \cite{Ramani:2015mcd}.

The availability of two-loop matrix elements for \ww production \cite{Gehrmann:2015ora, Caola:2015ila} has made possible NNLO predictions for the $W^+W^-$ production total cross section \cite{Gehrmann:2014fva} and differential distributions \cite{Grazzini:2016ctr}.
Furthermore, it is possible to extend both 
transverse momentum resummation\cite{Grazzini:2015wpa} to NNLL +NNLO (equivalent to NNLL'+NNLO in our log-counting convention) and jet-veto resummation calculations to  partial N$^3$LL+NNLO. The latter is carried out in this paper using soft collinear effective theory \cite{Bauer:2000ew, Bauer:2000yr, Bauer:2001ct, Bauer:2001yt, Beneke:2002ph, Beneke:2002ni, Hill:2002vw}.  (We include in our partial N$^3$LL resummation only the logarithmic contributions to the beam functions at two-loop.) This leads to more precise results with smaller scale uncertainties as compared to the NLO matched to NNLL jet-veto resummed results\cite{Jaiswal:2014yba,Becher:2014aya} which subsequently results in better tests of experiment-theory agreement.  We also include the loop-induced $gg$ initial state (which starts
at $\alpha_s^2$) \cite{Binoth:2006mf,Glover:1988fe, Dicus:1987dj} with the jet-veto resummation performed at NLL+LO.  The NLO contribution to the $gg$ initial
state is known\cite{Caola:2015rqy}, but not needed at the order we are working. Our numerical results are new, and provide an important test of the structure of the electroweak theory.

The paper is organized as follows. In Section \ref{sec:FO} we review the implementation of differential NNLO diboson production in Sherpa \cite{Gleisberg:2008ta}, and in Section \ref{sec:basics} the SCET formalism for jet veto resummation. In Section \ref{sec:results} we present our numerical results, along with a comparison with experimental results,  followed by concluding remarks.

\section{fixed order}
\label{sec:FO}
\makeatletter{}We make NNLO predictions for on-shell $W^+ W^-$ production with arbitrary cuts, using an implementation of $q_T$ subtraction \cite{Catani:2007vq, Catani:2009sm} in Sherpa \cite{Gleisberg:2008ta} with matrix element generators AMEGIC++ \cite{Krauss:2001iv} and Comix \cite{Gleisberg:2008fv}. We use the SCET transverse momentum factorization formalism \cite{Becher:2010tm}, with the 2-loop transverse parton distribution functions calculated in \cite{Gehrmann:2014yya}, to predict 
the NNLO cross section below a given $q_T^{\rm cut}$ for the \ww
pair momentum. This is implemented as a $K$ factor multiplying the Born cross section in Sherpa.
The NNLO cross section above the $q_T$ cut corresponds to an NLO cross section 
for $W^+ W^- j$ production \cite{Dittmaier:2007th,Melia:2012zg,Campbell:2015hya}, obtained using Sherpa's built-in 
NLO capability based on Catani-Seymour dipole subtraction \cite{Catani:1996vz} and the OpenLoops implementation \cite{Cascioli:2011va} of one-loop virtual matrix elements (a customized version with $n_f=4$ flavors is used). We extract the two-loop virtual correction from the public code {\tt qqvvamp} \cite{Gehrmann:2015ora} for diboson production, together with appropriate coupling factors specific to the \ww process. 
In Ref.~\cite{Gehrmann:2015ora}, Eq.\ (6.2)-(6.5) defines the IR-divergent, UV renormalized amplitude in the standard $\overline{\rm MS}$ scheme at the renormalization scale $\mu_r=m_{WW}$,
\begin{equation}
\Omega (\epsilon)= \Omega^{(0)} + \left( \frac{\alpha_s}{2\pi} \right) \Omega^{(1)} + \left( \frac{\alpha_s}{2\pi} \right)^2 \Omega^{(2)}.
\label{eq:QCDamp}
\end{equation}
Eq.\ (6.6) of Ref.\ \cite{Gehrmann:2015ora} defines the IR-finite amplitude in the scheme of \cite{Catani:2013tia}, and can be re-written as
\begin{equation}
\Omega^{\text{finite}} = \Omega(\epsilon) \cdot I (\epsilon),
\end{equation}
where
\begin{equation}
I (\epsilon) = 1 - \left( \frac{\alpha_s}{2\pi} \right) I_1 (\epsilon) - \left( \frac{\alpha_s}{2\pi} \right)^2 I_2 (\epsilon).
\end{equation}

Several checks are performed to confirm the validity of our extraction of virtual amplitudes from the {\tt qqvvamp} library. First, when only Drell-Yan type diagrams are taken into account by the code, we reproduce the well-known 1-loop and 2-loop hard functions for the 
Drell-Yan process \cite{Anastasiou:2003yy,Catani:2009sm}. Second, we are able to reproduce known 1- and 2-loop diphoton production virtual corrections \cite{Anastasiou:2002zn} from {\tt qqvvamp}, by setting the outgoing boson mass close to zero. Third, the one-loop $W^+ W^-$ virtual correction extracted from the library agrees with the known results in \cite{Frixione:1993yp}.

We adopt the $n_f=4$ scheme in both the fixed order and resummed calculations, 
to avoid complications from top resonances.  In the $n_f=5$ scheme, effects from the top resonance first enter the \ww rate at NLO\cite{Gehrmann:2014fva}.
We use MSTW 2008 NNLO $n_f=4$ parton distributions \cite{Martin:2010db}.
To speed up the numerical evaluation, we fit the 2-loop virtual correction for both $d$-type and $u$-type quarks,
as a function of the pair invariant mass $m_{WW}$ and the polar angle $\theta$. We use cubic interpolation on a two-dimensional grid of size $34 \times 89$, achieving an accuracy of $10^{-4}$ in almost all phase space points. We use the physical constants $G_F=1.1663787\times 10^{-5}$ GeV$^{-2}$, $m_W=80.399$ GeV, and $m_Z = 91.1876$ GeV.

Though we will present  cross sections  with kinematic cuts in latter parts of the paper, we first check that the total cross section from our implementation reproduces known results, at the renormalization scale and factorization scale $\mu_r=\mu_f=m_W$. For the 13 TeV $p p \rightarrow W^+ W^-$ total cross section, we obtain $(118.4 \pm 0.4)$ pb with $q_T^{\rm cut} = m_{WW} / 200$ and 46 million integration points, or $(118.8 \pm 0.8)$ pb with $q_T^{\rm cut} = m_{WW} / 2000$ and 342 million integration points, demonstrating excellent $q_T^{\rm cut}$ independence through one order of magnitude. Both of the above results agree with the known NNLO results in \cite{Gehrmann:2014fva} within statistical errors. For the rest of this paper, we will use $q_T^{\rm cut} = m_{WW} / 200$, and omit the uncertainty from $q_T^{\rm cut}$ dependence, estimated to be around $0.34\%$ from the two numbers above.
For 8 TeV, we obtain $(59.94\pm 0.16)$ pb with $q_T^{\rm cut} = m_{WW} / 200$, again in close agreement with Ref.\ \cite{Gehrmann:2014fva}. For the purpose of comparison, these results include the $gg$ box diagram (without the small interference effect from Higgs intermediate states), which contributes $3.9$ pb at 13 TeV and $1.5$ pb at 8 TeV.

\section{Jet veto resummation}
\label{sec:basics}
\makeatletter{}We study the process $q \bar{q} (gg) \rightarrow W^+ W^- + X$ where $X$ are hadronic jets 
satisfying the jet-veto condition,  $p_T^{\textrm{jet}} < \ptv$. The core Born process $q \bar q \rightarrow W^+ W^-$ starts at $\mathcal O(\alpha_s^0)$, while the core Born process $gg \rightarrow W^+ W^-$ is loop-induced and starts at $\mathcal O(\alpha_s^2)$. Notice that the resummed corrections to $q\bar q \rightarrow W^+ W^-$ also include $gg$ initial state contributions, in e.g.\ the $\mathcal O(\alpha_s^2)$ order double-real part, but this is distinct from the loop-induced $gg \rightarrow W^+ W^-$ which may be considered as a separate process. Due to the presence of multiple scales, large logarithms of the form $\alpha_s^n\ln^m\lambda$ 
($m \le 2n$) arise at higher orders in perturbation theory, with
$\lambda \equiv \left({\ptv\over m_{WW}}\right) \ll 1$ being the ratio of two scales. Methods for resumming these large logarithms are developed in Refs. \cite{Banfi:2012yh, Banfi:2012jm, Berger:2010xi, Tackmann:2012bt, Stewart:2013faa, Becher:2012qa, Becher:2013xia}.
We employ SCET formalism with $\lambda$ as 
the power-counting parameter of the effective theory. While we work at leading order in SCET, 
the expansion in $\alpha_s$  in the context of resummation has differing conventions in the literature 
that we wish to clarify. Counting large logarithms $\log \lambda \sim \alpha_s^{-1}$, 
we define N$^{n+1}$LL resummation as the expansion in $\alpha_s$ up to and including 
$\co(\alpha_s^{n})$ terms. 

Using the  SCET formalism with an analytic regulator to deal with rapidity divergences, the jet-veto cross-section factorizes as \cite{Becher:2012qa, Jaiswal:2015nka}, 
\begin{eqnarray}
&& {d\sigma\over dm_{WW} \, dy \, d \cos\theta}
= \, \sum_{i,j=q, {\overline q},g}\ch_{ij}(m_{WW}, \mu_f, \mu_h, \cos\theta)
\nonumber \\
&& \qquad \qquad \times \,
\cb_i(\xi,\ptv,\mu_f,\nu_B, R) \,   \cb_j(\bar{\xi}, \ptv,\mu_f, \bar{\nu}_B, R) \nonumber \\
&& \qquad \qquad \times \, 
{\left( \frac{\nu_S \bar{\nu}_S}{\nu_B \bar{\nu}_B} \right)}^{g(\mu_f)} \cs (\ptv, \mu_f, \nu_S, \bar{\nu}_S, R) \, ,
\label{factorization}
\end{eqnarray}
where $\ch$, $\cb$ and $\cs$ are the hard, beam and soft 
functions, respectively, $R$ is the jet radius and first enters at two-loops, and 
$\nu_S , \bar{\nu}_S, \nu_B, \bar{\nu}_B $ are rapidity renormalization scales.
The function $g(\mu_f)$ is defined in Eq. \ref{gdef}.  In principle, the beam function could depend on an independent scale, $\mu_b$, but
we always take $\mu_b=\mu_f$. 
 We ignore factorization-violating effects which start at $\mathcal O(\alpha_s^4)$ \cite{Gaunt:2014ska, Zeng:2015iba, Rothstein:2016bsq}.

\subsection{Hard function}
The hard function $\ch$ is roughly speaking the square of the SCET Wilson coefficient $C$ which is obtained by matching SCET to QCD 
at a hard scale, $\mu_h \sim m_{WW}$,  and then by renormalization group (RG) evolving down to the factorization 
scale, $\mu_f \sim \ptv$.

For the NLL+LO resummation of $gg \rightarrow W^+ W^-$, we only need the LO hard function, which may be easily extracted from the Born-level distributions generated by Sherpa. For the rest of this subsection, we will focus on the 2-loop hard function for $q \bar q \rightarrow W^+ W^-$, needed for N$^3$LL resummation.

The SCET hard matching coefficient $C$ is given by,
\begin{equation}
C = \Omega (\epsilon) / \Omega_{sc} (\epsilon) = \Omega(\epsilon) \cdot I^{\rm SCET} (\epsilon),
\label{eq:scetC}
\end{equation}
where,
\begin{equation}
I^{\rm SCET} (\epsilon) = 1 - \left( \frac{\alpha_s}{2\pi} \right) I_1^{\rm SCET} (\epsilon) - \left( \frac{\alpha_s}{2\pi} \right)^2 I_2^{\rm SCET} (\epsilon).
\end{equation}
In Eq.\ \eqref{eq:scetC}, $\Omega$ is the full QCD amplitude defined in Eq.\ \eqref{eq:QCDamp}, and $\Omega_{sc}$ is the amplitude obtained from the SCET Lagrangian by setting the hard matching coefficient to $1$. For massless parton scattering, $\Omega_{sc}$ is a dimensionless integral at every non-zero loop order, and contains pure IR poles after $\overline{\rm MS}$ UV subtraction (here ``pure poles'' means there are no $\epsilon$-independent constant terms).

In other words, the main difference between the definition of the SCET hard function and the IR subtraction scheme used in {\tt qqvvamp} is the fact that SCET uses pure poles for subtraction. The amplitude $\Omega^{\rm finite}$ from {\tt qqvvamp} can be squared and converted into the SCET hard function $\ch$ at the central hard scale $\mu_h = m_{WW}$ by
using the relation, 
\begin{align}
\tilde{\ch} (\mu_h=m_{WW})
&= \left| \Omega^{\rm SCET} \right|^2 \nonumber \\
&= \left| \Omega^{\rm finite} \right|^2 \left[ 1 + \left( \frac{\alpha_s}{2\pi} \right) \hat I_1 (\epsilon) + \left( \frac{\alpha_s}{2\pi} \right)^2 \hat I_2 (\epsilon) \right],\label{eq:hardCentral}
\end{align}
where,
\begin{align}
\hat I_1 & = \frac {\pi^2 C_F} {6}, \nonumber \\
\hat I_2 & = \frac 1 {72} C_F \left( \pi^4 C_F + 36 \delta_{q_T}^{(1)} + 12 \pi^2 K + 48 \beta_0 C_F \zeta_3 \right) ,
\end{align}
and $\delta_{q_T}^{(1)}$ and $K$ are defined in Eq. (6.11) of Ref.\ \cite{Gehrmann:2015ora}. Once we have the 1-loop and 2-loop hard function $\tilde{\ch}$ at the scale $m_{WW}$, we use SCET RG running to restore the full dependence on the scale $\mu_h$. Writing
\begin{equation}
\ch =\ch^{(0)} + \ch^{(1)} a_s + \ch^{(2)} a_s^2,
\end{equation}
where $a_s = \alpha_s / (4 \pi)$, and
\begin{equation}
L_h = \log \frac {\mu_h}{m_{WW}},
\end{equation}
the $\mu_h$ dependence of $\ch^{(2)}$ is
\begin{align}
\ch^{(2)} (\mu_h) &= \ch^{(2)} (m_{WW}) + 2 \ch^{(1)} (m_{WW}) \gamma_0 L + 2 \gamma_1 L \nonumber \\
&\quad - 2 \Gamma_1 L^2 - 2 \ch^{(1)} (m_{WW}) \Gamma_0 L^2 + 2 \gamma_0^2 L^2 \nonumber \\
&\quad -4 \Gamma_0 \gamma_0 L^3 + 2 \Gamma_0^2 L^4 + 2 \beta_0
\left( \ch^{(1)} (m_{WW}) L + \gamma_0 L^2 - \frac 2 3 \Gamma_0 L^3  \right),
\end{align}
where $\Gamma_n$ and $\gamma_n$ are the perturbative expansions of the cusp and non-cusp part of the anomalous dimension for the Drell-Yan like hard function, found in e.g.\ \cite{Becher:2007ty}.

The SCET Wilson coefficients $C$ satisfy the RG equation,
\beq
\mu \frac{\der C(\mu)}{\der \mu} = \left(\Gamma^{\rm{cusp}}_i  \log \left( \frac{-m_{WW}^2 - i \epsilon}{\mu_h^2} \right) 
+ 2 \gamma_i \right)  C(\mu)\, ,
\eeq
where $\gamma$ and $\Gamma^{\rm{cusp}}$ are the  anomalous and cusp-anomalous dimensions, respectively, with $i=F$ for a $q \bar{q}$ initiated process, while $i=A$ for a $gg$ initiated-process.  
For brevity, we have suppressed Lorentz indices as well as external particle momentum dependence in the Wilson coefficients.
N$^3$LL accuracy requires 3-loop anomalous dimensions which are known and 4-loop cusp anomalous dimensions for which we use the Pade approximation. Up to 3-loops, the gluon 
anomalous dimensions can be obtained from the quark anomalous dimensions by replacing $C_F$ with $C_A$. 

The solution to the RG evolution of the SCET Wilson coefficients can be written as \cite{Becher:2007ty},
\beq
C(\mu) = \cu(\mu, \mu_h) C(\mu_h)\, ,
\label{WilsonC}
\eeq
where the evolution function $\cu$ is given by, 
\begin{eqnarray}
\cu(\mu, \mu_h) &=  \exp & \Big[ 2 S(\mu,\mu_h) - 2 a_\gamma(\mu,\mu_h)   \nonumber \\
 & & -a_\Gamma(\mu,\mu_h) \log \left( \frac{-m_{WW}^2 - i \epsilon}{\mu_h^2} \right) \Big]
\end{eqnarray}
and expressions for $S$, $a_\gamma$ and $a_\Gamma$ can be found in Ref.  \cite{Becher:2007ty}.

Using Eq.\eqref{WilsonC}, we get the following result for the resummed hard function
\beq
\ch(\mu, \mu_h) = \left|  \cu(\mu, \mu_h) \right|^2 \tilde{\ch}(\mu_h)
\eeq

\subsection{Beam function}

The beam functions, at lowest order are simply the 
PDFs, $\cb_i = \phi_i$, while at higher orders the PDFs are convoluted with kernels as follows,  
\begin{equation}
\cb_i(\xi,\ptv,\mu,R)=\int_{\xi}^1{dz\over z}\sum_{k}I_{i\leftarrow k}(z,\ptv,\mu,R)
\phi_k({\xi\over z},\mu)\, .
\end{equation}
Expanding the kernels in $\alpha_s$, we have
\beq
I_{i\leftarrow k}(z,\ptv,\mu,R) = \hat{I}_{q i}^{(0)} + a_s \hat{I}_{q i}^{(1)} + a_s^2 \hat{I}_{q i}^{(2)} + \cdots
\eeq
where $a_s \equiv \alpha_s(\mu)/(4 \pi)$. The kernels $\hat{I}$ are in general functions of $\{z, \mu, \ptv, R\}$.
The $\mathcal O(\alpha_s)$ kernels are
\begin{align}
\hat{I}_{q q}^{(1)} &= - \left[ \left(\frac{d_1}{2} +  \gamma_0 \right) L_\perp 
+ \Gamma_0 \frac{L_\perp^2}{4} \right] \delta(1-z) \nonumber \\
&\quad - \cp_{qq}^{(1)} \frac{L_\perp}{2} + \cR_{q q}^{(1)} (z), \\
\hat{I}_{q  g}^{(1)} &= - \cp_{qg}^{(1)} \frac{L_\perp}{2} + \cR_{q g}^{(1)} (z),
\end{align}
where we define $L_{\perp} = 2 \log ( \mu / p_T^{\rm veto} )$, $\gamma_n$ and $\Gamma_n$ denote the non-cusp and cusp anomalous dimensions for the hard function at order $n$, and $\cp_{ij}^{(n)}$ denotes the DGLAP splitting kernel at order $n$. The non-logarithmic terms in the above equations, denoted by $\cR_{ij}^{(n)}$, can be deduced from Ref. \cite{Becher:2010tm}, and are given by
\begin{align}
\cR_{q q}^{(1)} (z) &= C_F \left[ 2 (1-z) - \frac{\pi^2}{6} \delta \left( 1-z \right) \right], \nonumber \\
\cR_{q g}^{(1)} (z) &= 4 T_F z (1-z).
\end{align}
In addition, we include approximate NNLO kernels by solving RG evolution equations to obtain the terms that depend on $L_{\perp}$.\footnote{Approximate NNLO beam functions have been previously obtained in a different rapidity regularization scheme in Ref.  \cite{Li:2014ria}.}.
Here, $d_1=0$ while $d_2$ depends on the jet radius $R$ and can be found in Ref. \cite{Becher:2013xia}. The  N$^3$LL ingredient $d_3$ also depends on $R$. The leading $\log R$ terms in $d_3$ are known and can be extracted from 
\cite{Dasgupta:2014yra, Banfi:2015pju}.  We include these terms in our numerical results.

The result for $\hat{I}_{qq}^{(2)}$, describing  the quark beam function corresponding to the quark PDF of the same flavor, is
\begin{align}
\hat{I}_{qq}^{(2)} &= \frac{(\Gamma_0)^2}{2} \left( \frac{L_\perp}{2}  \right)^4 \nonumber \\
 		     & \quad  + \left[ (d_1- \frac 4 3 \beta_0 + 2 \gamma_0) \delta(1-z)  \cp_{qq}^{(1)}(z) \right]\Gamma_0  \left( \frac{L_\perp}{2}  \right)^3 \nonumber  \\
		    &  \quad  + \Bigg\{ \left[ \frac{d_1}{2} (d_1 + 4 \gamma_0) 
		             - 2\beta_0 (d_1 + \gamma_0) + 2\gamma_0^2 - \Gamma_1 \right]\nonumber \\
			& \quad \times \delta(1-z) - \Gamma_0 \cR_{qq}^{(1)}(z) + (d_1 - \beta_0 + 2 \gamma_0) \cp_{qq}^{(1)}(z) \nonumber \\
			& \quad + \frac{\cp_{qq}^{(1)}(z) \otimes \cp_{qq}^{(1)}(z)}{2}           
		    + \frac{\cp_{qg}^{(1)}(z) \otimes \cp_{gq}^{(1)}(z)}{2} \Bigg\} \left( \frac{L_\perp}{2}  \right)^2 \nonumber \\
		    & \quad  + \Bigg\{- (d_2 + 2 \gamma_1) \delta(1-z) 
		              - (d_1 - 2 \beta_0 + 2 \gamma_0)\cR_{qq}^{(1)}  \nonumber  \\
		    &  \quad - \cp_{qq}^{(2)}(z) 
		           - \cR_{qq}^{(1)}(z) \otimes \cp_{qq}^{(1)}(z) \nonumber \\
		    & \quad - \cR_{qg}^{(1)}(z) \otimes \cp_{gq}^{(1)}(z) \Bigg\}
		           \left( \frac{L_\perp}{2}  \right)^1 +  \cR_{qq}^{(2)}(z) \left( \frac{L_\perp}{2}  \right)^0\, ,
\end{align}
where,
\beq
f(z, \dots) \otimes g (\xi, \dots) =  \int_\xi^1 \frac{\der z}{z} f(z, \dots) g (\xi/z, \dots)\, .
\eeq

The result for $\hat{I}_{qg}^{(2)}$, describing the quark/anti-quark beam function involving  the gluon PDF, is
\begin{align}
\hat{I}_{qg}^{(2)} &= \Gamma_0 \cp_{qg}^{(1)}(z) \left( \frac{L_\perp}{2}  \right)^3 \nonumber \\
           &  \quad + \Bigg\{ -\Gamma_0 \cR_{qg}^{(1)}(z) + (d_1 - \beta_0 + 2\gamma_0)\cp_{qg}^{(1)}(z) \nonumber \\
           &  \quad + \frac{\cp_{qq}^{(1)}(z) \otimes \cp_{qg}^{(1)}(z)}{2} + \frac{\cp_{qg}^{(1)}(z) \otimes \cp_{gg}^{(1)}(z)}{2}
                 \Bigg\} \left( \frac{L_\perp}{2}  \right)^2 \nonumber \\
           &  \quad + \Bigg\{ -(d_1 - 2\beta_0 + 2 \gamma_0)\cR_{qg}^{(1)}(z) -\cp_{qg}^{(2)}(z) \nonumber \\
           &  \quad - \cR_{qq}^{(1)}(z) \otimes \cp_{qg}^{(1)}(z) - \cR_{qg}^{(1)}(z) \otimes \cp_{gg}^{(1)}(z) \Bigg\} \left( \frac{L_\perp}{2}  \right)^1 \nonumber \\	
           &  \quad + \cR_{qg}^{(2)}(z) \left( \frac{L_\perp}{2}  \right)^0.
\end{align}
Finally, the result for $\hat{I}_{qq'}^{(2)}$, describing the quark beam function corresponding to the PDF of an anti-quark or a quark of a different flavor, is
\begin{align}
\hat{I}_{qq'}^{(2)} &= \Bigg\{ \frac{\cp_{qg}^{(1)}(z) \otimes \cp_{gg'}^{(1)}(z)}{2}
                 \Bigg\} \left( \frac{L_\perp}{2}  \right)^2 \nonumber \\
           & \quad + \Bigg\{-\cp_{qq'}^{(2)}(z) - \cR_{qg}^{(1)}(z) \otimes \cp_{gq'}^{(1)}(z) \Bigg\} \left( \frac{L_\perp}{2}  \right)^1 \nonumber  \\	
           & \quad + \cR_{qq'}^{(2)}(z) \left( \frac{L_\perp}{2}  \right)^0.
\end{align}
The coefficients of $(L_\perp /2 )^0$ in the above three equations are unknown, and are set to zero in our calculation. Due to the missing non-logarithmic 
contributions in the beam functions, we can not claim full N$^3$LL accuracy. We term our resummation with approximate two-loop beam functions as partial N$^3$LL, abbreviated as N$^3$LL$_p$.

\subsection{Rapidity Renormalization Group}
We adopt a regularization scheme for rapidity divergences where the soft-function $\cs = 1$ to all orders in perturbation theory. 
In the language of the rapidity renormalization group  \cite{Chiu:2011qc,Chiu:2012ir}, the factor  
$(\nu_S \bar{\nu}_S/ \nu_B \bar{\nu}_B)^{g(\mu)}$ in Eq.\  \eqref{factorization}  is the result of the RG evolution of the rapidity scale $\nu$ 
between soft modes ($\nu_S \sim \bar{\nu}_S \sim \mu_f$) and collinear modes ($\nu_B \sim \bar{\nu}_B \sim m_{WW}$). 
Suppressing the  $\ptv$, $R$ and $\xi$ dependence of the beam functions for brevity, the rapidity scale variation around their central values 
at the boundaries of  the rapidity RG can be summarized as \cite{Jaiswal:2015nka}, 
\begin{eqnarray}
&& {\left( \frac{\nu_S \bar{\nu}_S}{\nu_B \bar{\nu}_B} \right)}^{g(\mu_f)}
\cb_i(\mu_f,\nu_B) \,   \cb_j(\mu_f, \bar{\nu}_B) \nonumber \\
&& = \left(r {\mu_f^2\over m_{WW}^2}\right)^{g(\mu_f)} 
\left[ r^{-g(\mu_f)} \cb_i(\mu_f) \,   \cb_j(\mu_f) \right]_{\co(\alpha_s^n)}
\label{eq:RRG}
\end{eqnarray}
where $1/2 < r < 2$ contributes to the rapidity scale uncertainty. The quantity inside the square bracket 
is expanded to $\co(\alpha_s^n)$ for N$^{n+1}$LL resummation while $g(\mu_f)$ outside the square bracket 
must be evaluated to  $\co(\alpha_s^{n+1})$. For the central scale $r=1$, Eq.\  \eqref{eq:RRG} reduces to a form 
given in Ref. \cite{Becher:2012qa} where the factor $(\mu_f^2/m_{WW}^2)^{g(\mu_f)}$  has been termed 
the `collinear anomaly'.

Writing $g(\mu)$ as a polynomial in $L_\perp \equiv 2 \log(\mu/\ptv)$, 
\beq
g(\mu) \equiv \sum\limits_{p=0}^\infty g^{(p)} L_\perp^p \, ,
\label{gdef}
\eeq
the perturbative expansions for $g^{(p)} $ are given by \cite{Jaiswal:2015nka},
\begin{eqnarray}
g^{(0)} &=& a_s d_1 + a_s^2 d_2 + a_s^3 d_3  \nonumber\\
g^{(1)} &=& a_s \Gamma_0 +a_s^2 (\beta_0 d_1 + \Gamma_1)   
+ a_s^3 (2 \beta_0 d_2 + \beta_1 d_1 + \Gamma_2) \nonumber\\
g^{(2)} &=& a_s^2  \frac{\beta_0 \Gamma_0}{2}   
+ a_s^3 (\beta_0^2 d_1 + \frac{\beta_1 \Gamma_0}{2} + \beta_0\Gamma_1)  \nonumber\\
g^{(3)} &=&  a_s^3 \frac{\beta_0^2 \Gamma_0}{3}  
\end{eqnarray}
where $\Gamma_n$ and $\beta_n$ denote the cusp anomalous dimension and the QCD beta function at order $n$.

The jet-veto resummed result can be combined with the fixed order result,
\begin{equation}
\sigma^{\rm matched}_{\text{N}^3\text{LL}_p+{\rm NNLO}}=\sigma^{\rm res}_{{\rm N}^3{\rm LL}_p}+\sigma_{\rm NNLO}- \sigma^{\rm res}_{{\rm N}^3{\rm LL}_p}|_{\rm exp.}\, ,
\end{equation}
and
$\sigma^{\rm res}|_{\rm exp.}$ is the resummed result expanded up to ${\cal{O}}(\alpha_s^2)$. For the loop-induced $gg$ contribution, we only match to LO, in which case the matching is trivial. These results can be directly compared with experiment for a given jet veto cut.

\section{Results}
\label{sec:results}
\makeatletter{}Applying the formalism described in the above sections, we compute fixed-order and resummed \ww cross section under a jet veto. 

We present the fixed order NLO and NNLO cross sections with a jet veto at $13$ TeV for two different central scale choices in Tab.\ \ref{tab:mubchoice}.  Our default choice in the remainder of this section is $\mu_f=\mu_r=\ptv$, which is convenient because we match resummation and fixed order results at the central scale $\mu_f = \ptv$.  We see, however, that if we had chosen a larger scale, say $\mu_f=\mu_r=2m_W$, the fixed order results would not be much changed. We will comment on the smallness of the scale dependence later. The scale uncertainties in the NLO and NNLO results are obtained by simultaneously varying $\mu_r$ and $\mu_f$ up and down by a factor of $2$ around the central value $\ptv$.

The central hard scale is taken to be $\mu_h^2=m_{WW}^2$ for the resummation of $q \bar q \rightarrow W^+ W^-$, but the time-like scale choice $\mu_h^2 = - m_{WW}^2$ is used for the loop-induced $gg$ channel to resum the $\pi^2$ terms that give large corrections to the LO hard function. The central factorization and beam function scales are $\mu_f=\mu_b= \ptv$, and the rapidity scale variation factor $r$ defined in Ref.\ \cite{Jaiswal:2015nka} is centered at $1$. We vary these scales up and down by a factor of 2 in order to estimate uncertainties from as yet uncomputed higher order contributions. As an example, the cross sections with these scale variations are tabulated in Table \ref{tab:t3} for $q\bar{q}$ processes with the CMS jet veto cut ($\ptv=30$ GeV, $R=0.5$) at 13 TeV.\footnote{The qualitative features of this table also hold true at 8 TeV and 
for the ATLAS jet veto cuts at both 8 and 13 TeV.} 

\begin{table}[tb]
  \begin{tabular} {| c | c | c | }
    \hline
     & ATLAS & CMS \\
     \hline
    NLO ($\mu_f=\mu_r=\ptv$) &  $71.18 \pm 2.17$ & $74.68 \pm 0.77$ \\
    \hline
     NLO ($\mu_f=\mu_r=2m_W$) & $72.82 \pm 1.69$ &  $77.26 \pm 1.30$ \\
     \hline
     NNLO ($\mu_f=\mu_r=\ptv$) & $ 70.01  \pm 1.09 $ &  $73.46 \pm 0.16$ \\
     \hline
     NNLO ($\mu_f=\mu_r=2m_W$) &  $69.68 \pm 0.88$ & $74.89 \pm 0.79$ \\ 
     \hline
  \end{tabular}
  \caption{Fixed order 13 TeV jet veto cross-sections for two disparate factorization/beam function central scale choices ($\ptv$ and $2m_W$) for the ATLAS configuration  $\ptv=25$ GeV and R=0.4 and CMS configuration $\ptv=30$ GeV and R=0.5 for on-shell \ww production. The scale variations are symmetrized, i.e.\ they are averages over the absolute values of the fluctuations at twice the central scale and half the central scale. All cross section numbers are listed in pb. These numbers do not include the $gg$ initial state.}
  \label{tab:mubchoice}
\end{table}

The rapidity scale variation at N$^3$LL$_p$+NNLO is artificially small at the central $\mu_b$, as shown in the last column of the 2nd-to-last row in Table \ref{tab:t3}, due to the lack of non-logarithmic terms in our approximate NNLO beam functions. We therefore estimate the rapidity scale variation fixing $\mu_h$ at the central scale, but $\mu_b$ at half the central scale.

\begin{figure}[tb]
\begin{centering}
\includegraphics[scale=0.4]{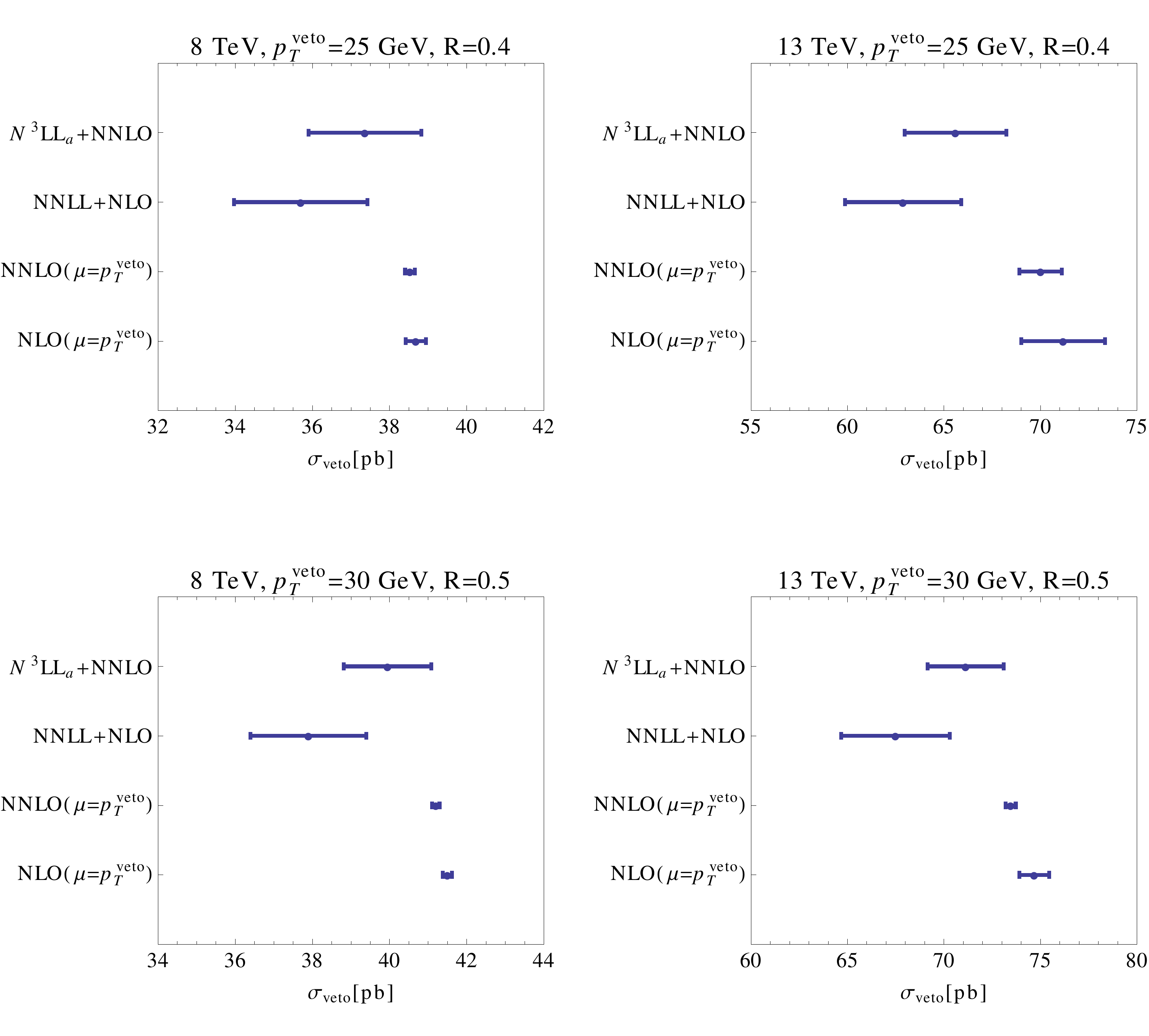}
\par\end{centering}
\caption{Summary of jet veto resummation results for $q \bar q \rightarrow W^+ W^-$. We include results at 8 TeV and 13 TeV, under ATLAS or CMS jet veto cuts.} 
  \label{fig:resultAll}
\end{figure}

The scale uncertainties from $\mu_h$, $\mu_b$ and $r$ variations are symmetrized for simplicity (i.e.\ we do not give asymmetric error bars), and added in quadrature to obtain the combined uncertainty. Due to the Monte Carlo integration used to compute the fixed order NNLO cross section, we also have an additional statistical error. However it is small compared to the combined scale variations.
\begin{table}[tb]
  \begin{tabular} {| c | c | c  |}
    \hline
    $\{ \mu_h / m_{WW}, \mu_b / \ptv ,r \}$ & NNLL+NLO &$\text{N}^3{\rm LL}_p$+NNLO  \\ 
    \hline
     $\sigma \{1,1,1\} $& $67.49$ & $71.18$ \\
    \hline
    \hline
   $\Delta \sigma \{1,(2,\frac{1}{2}),1\} $&  $0.93 $ & $1.58$ \\
    \hline
   
  $\Delta \sigma  \{(2,\frac{1}{2}),1,1\} $ & $1.92 $ &  $0.84 $\\
     \hline
 
$\Delta \sigma \{1,1,(2,\frac{1}{2})\} $ & $ 1.31 $ &  $ 0.04 $\\
     \hline 
 $\Delta \sigma    \{1,\frac{1}{2},(2,\frac{1}{2})\} $ &  $ 1.83 $ &  $ 0.79 $\\
     \hline
   
  \end{tabular}
  \caption{8 and 13 TeV scale variations for the CMS configuration $\ptv=30$ GeV and R=0.5 for  on-shell \ww production. The scale variations are symmetrized, i.e.\ they are averages over the absolute values of the fluctuations at twice the central scale and half the central scale. All cross section numbers are listed in pb. These numbers do not include the $gg$ initial state.}
  \label{tab:t3}
\end{table}
We present the final results most relevant to the experiments, R=0.4, $\ptv=25$ GeV for ATLAS in Table \ref{tab:t1} and  R=0.5, $\ptv=30$ GeV for CMS in Table \ref{tab:t2}  with the theoretical error bars. The information from these two tables is also summarized in the plot, Figure \ref{fig:resultAll}. We note the effects of the $\pi^2$ resummation by presenting
our final results both with and without this resummation (the last 2 lines in Tabs.~\ref{tab:t1}, and ~\ref{tab:t2}).  This effect can be seen to be small.
 \begin{table}[tb]
  \begin{tabular} {| c | c | c  |}
    \hline
    Order & 8 TeV  & 13 TeV \\ 
    \hline
     NLO & $38.68 \pm 0.26$ & $71.18 \pm 2.17$ \\
    \hline
    NNLL+NLO &  $35.70 \pm 1.73 $ & $62.88 \pm 3.01$ \\
    \hline
    NNLO ($\mu=\ptv$) & $ 38.53  \pm 0.09 \pm 0.09 $ &  $ 70.01  \pm 1.04  \pm 0.34 $\\
     \hline
     $\text{N$^3$LL}_p$+NNLO & $37.36 \pm 1.46 \pm 0.09$  &  $65.60 \pm 2.61 \pm 0.34$ \\
     \hline
     $gg$ NLL & $0.35 \pm 0.11$  &  $0.80 \pm 0.12$ \\
       \hline
     $gg$ NLL (with $\pi^2$)& $0.57 \pm 0.21$  &  $1.31 \pm 0.43$ \\
     \hline\hline
         Total $\text{N$^3$LL}_p$+NNLO +$gg$ NLL &$37.45\pm 1.47$&$66.40\pm 2.63$\\
     \hline
     Total $\text{N$^3$LL}_p$+NNLO +$gg$ NLL  (with $\pi^2$)&$37.93\pm 1.48$&$66.91\pm 2.67$\\
     \hline
  \end{tabular}
  \caption{8 and 13 TeV Jet veto cross sections for the ATLAS configuration $\ptv=25$ GeV and R=0.4
  for on-shell \ww production. All cross section numbers are listed in pb. The first error bar following each number denotes the scale uncertainty, while the 2nd error bar, if present, denotes the statistical error from Monte Carlo integration 
    in the implementation of NNLO in Sherpa with the $q_T$ subtraction cutoff $M_{WW}/200$.}
  \label{tab:t1}
\end{table}

Going to one higher order, i.e. NNLL+NLO $\rightarrow$ N$^3$LL$_p$ +NNLO, has an appreciable effect on the jet veto cross section prediction, due to both the higher accuracy in the resummation and the matching to higher fixed order calculations.  We observe that there is a reduction in scale uncertainties. We also present the corresponding fixed order results in order to illustrate the effects of resummation. One notices that the scale dependence of  the corresponding fixed order calculations are much smaller than those of the  resummed results, and in fact much smaller (as a percentage) than the uncertainties in the $W^+ W^-$ total cross section reported in Ref.\ \cite{Gehrmann:2014fva}. This is because of the under-estimation of the scale dependence due to an accidental cancellation, as observed in Ref.\ \cite{Becher:2012qa}.

\begin{table}[tb]
  \begin{tabular} {| c | c | c  |}
    \hline
    Order & 8 TeV  & 13 TeV \\ 
    \hline
     NLO &  $41.50 \pm 0.12$  & $74.68 \pm 0.77$ \\
    \hline
    NNLL+NLO &  $37.90 \pm 1.50 $ &$67.49 \pm 2.81$\\
    \hline
    NNLO ($\mu=\ptv$)& $41.20 \pm 0.03  \pm 0.09$ &  $73.46 \pm 0.19  \pm 0.18$\\
     \hline
     $\text{N$^3$LL}_\text{a}$+NNLO & $39.95 \pm 1.13 \pm 0.09$  &  $71.13 \pm 1.96 \pm 0.18$ \\
     \hline
     $gg$ NLL & $0.43 \pm 0.37$  &  $1.01 \pm 0.29$ \\
     \hline
     $gg$ NLL (with $\pi^2$)& $0.71 \pm 0.25$  &  $1.65 \pm 0.54$ \\
     \hline\hline
     Total $\text{N$^3$LL}_p$+NNLO +$gg$ NLL &$40.38\pm1.19$&$71.42\pm 1.99$\\
     \hline
       Total $\text{N$^3$LL}_p$+NNLO +$gg$ NLL  (with $\pi^2$)&$40.66\pm1.16$&$72.78\pm 2.04$\\
     \hline
  \end{tabular}
  \caption{Same as Table \ref{tab:t1}\,, but for the CMS configuration $\ptv=30$ GeV and R=0.5 for on-shell \ww production. All cross section numbers are listed in pb.}
  \label{tab:t2}
\end{table}

Finally we compare the total $q\bar{q}$+gg cross-section under a jet-veto with full luminosity Run-1 experimental results from the LHC. For CMS, the unfolded result, available in Table 6 of \cite{Khachatryan:2015sga}, is $44.0 \pm 0.7 \text{(stat)} \pm 2.5 \text{(exp)} \pm 1.4 \text{(theo)} \pm 1.1 \text{(lumi)}$, in picobarns. This is in good agreement with our theoretical prediction of $(40.66 \pm 1.16)$ pb. For ATLAS, our theoretical prediction is $(37.93\pm 1.48)$ pb, which may be compared with experimental data once an unfolded jet vetoed cross section is produced by ATLAS.

\section{Conclusions}
We performed partial N$^3$LL$_a$+NNLO resummation for on-shell $W^+ W^-$ production at the LHC using soft collinear effective theory. We relied on the publicly available two-loop virtual corrections, as well as the capability of the Sherpa framework for computing differential NNLO diboson production cross sections via $q_T$ subtraction. The small loop-induced $gg$ contribution has been resummed to NLL+LO, including $\pi^2$ resummation to simulate the large K factor corrections to the LO cross section. Our predicted fiducial cross sections show higher central-scale values than the NNLL+NLO results, with reduced scale uncertainties which albeit overlap with the uncertainty band of NNLL+NLO. Our results show  qualitative agreement with Ref.  \cite{Grazzini:2015wpa} which simulated jet veto effects using transverse momentum resummation matched to NNLO results for $W^+ W^-$ production. In particular, we confirm that resummation reduces the jet veto efficiency compared with NNLO, and our results show percentage errors that are comparable with Ref.\ \cite{Grazzini:2015wpa}. However, our resummation formalism also takes the jet algorithm and jet radius dependence into account.  The gg contribution is included at NLL to obtain the total jet-veto cross section for the $W^+ W^-$  process.   The final result is in good agreement with experimental results. A possible future direction would be matching parton shower to NNLO for diboson production in Sherpa, as has been done for Drell-Yan and Higgs production \cite{Hoeche:2014aia, Hoche:2014dla}, and compare with the analytic resummation results of this paper.

\section*{Acknowledgements}
We would like to thank Stefan Hoeche for help with implementing NNLO $q_T$ subtraction in Sherpa and for many useful discussions about the Sherpa framework. We also thank Stefano Pozzorini and Philipp Maierhöfer for providing a customized OpenLoops library with $n_f=4$ for the real-virtual part of NNLO $WW$ production. The work of SD is supported by the U.S. Department of Energy under contract DE-AC02-76SF00515.
The work of HR is supported in part by NSF
CAREER Award NSF-PHY-1056833.  The work  of MZ is supported by the Department of Energy under Award Number DE-{S}C0009937.
 Fermilab is operated by Fermi Research Alliance,
LLC under Contract Number DE-AC02-07CH11359 with the United States Department of Energy. 

\bibliography{paper}

\begin{thebibliography}{75}
\expandafter\ifx\csname natexlab\endcsname\relax\def\natexlab#1{#1}\fi
\expandafter\ifx\csname bibnamefont\endcsname\relax
  \def\bibnamefont#1{#1}\fi
\expandafter\ifx\csname bibfnamefont\endcsname\relax
  \def\bibfnamefont#1{#1}\fi
\expandafter\ifx\csname citenamefont\endcsname\relax
  \def\citenamefont#1{#1}\fi
\expandafter\ifx\csname url\endcsname\relax
  \def\url#1{\texttt{#1}}\fi
\expandafter\ifx\csname urlprefix\endcsname\relax\def\urlprefix{URL }\fi
\providecommand{\bibinfo}[2]{#2}
\providecommand{\eprint}[2][]{\url{#2}}

\bibitem[{\citenamefont{Gehrmann
  et~al.}(2014{\natexlab{a}})\citenamefont{Gehrmann, Grazzini, Kallweit,
  Maierh{\"o}fer, von Manteuffel, Pozzorini, Rathlev, and
  Tancredi}}]{Gehrmann:2014fva}
\bibinfo{author}{\bibfnamefont{T.}~\bibnamefont{Gehrmann}},
  \bibinfo{author}{\bibfnamefont{M.}~\bibnamefont{Grazzini}},
  \bibinfo{author}{\bibfnamefont{S.}~\bibnamefont{Kallweit}},
  \bibinfo{author}{\bibfnamefont{P.}~\bibnamefont{Maierh{\"o}fer}},
  \bibinfo{author}{\bibfnamefont{A.}~\bibnamefont{von Manteuffel}},
  \bibinfo{author}{\bibfnamefont{S.}~\bibnamefont{Pozzorini}},
  \bibinfo{author}{\bibfnamefont{D.}~\bibnamefont{Rathlev}}, \bibnamefont{and}
  \bibinfo{author}{\bibfnamefont{L.}~\bibnamefont{Tancredi}},
  \bibinfo{journal}{Phys. Rev. Lett.} \textbf{\bibinfo{volume}{113}},
  \bibinfo{pages}{212001} (\bibinfo{year}{2014}{\natexlab{a}}),
  \eprint{1408.5243}.

\bibitem[{\citenamefont{Grazzini et~al.}(2016)\citenamefont{Grazzini, Kallweit,
  Pozzorini, Rathlev, and Wiesemann}}]{Grazzini:2016ctr}
\bibinfo{author}{\bibfnamefont{M.}~\bibnamefont{Grazzini}},
  \bibinfo{author}{\bibfnamefont{S.}~\bibnamefont{Kallweit}},
  \bibinfo{author}{\bibfnamefont{S.}~\bibnamefont{Pozzorini}},
  \bibinfo{author}{\bibfnamefont{D.}~\bibnamefont{Rathlev}}, \bibnamefont{and}
  \bibinfo{author}{\bibfnamefont{M.}~\bibnamefont{Wiesemann}}
  (\bibinfo{year}{2016}), \eprint{1605.02716}.

\bibitem[{\citenamefont{Catani et~al.}(2009)\citenamefont{Catani, Cieri,
  Ferrera, de~Florian, and Grazzini}}]{Catani:2009sm}
\bibinfo{author}{\bibfnamefont{S.}~\bibnamefont{Catani}},
  \bibinfo{author}{\bibfnamefont{L.}~\bibnamefont{Cieri}},
  \bibinfo{author}{\bibfnamefont{G.}~\bibnamefont{Ferrera}},
  \bibinfo{author}{\bibfnamefont{D.}~\bibnamefont{de~Florian}},
  \bibnamefont{and} \bibinfo{author}{\bibfnamefont{M.}~\bibnamefont{Grazzini}},
  \bibinfo{journal}{Phys. Rev. Lett.} \textbf{\bibinfo{volume}{103}},
  \bibinfo{pages}{082001} (\bibinfo{year}{2009}), \eprint{0903.2120}.

\bibitem[{\citenamefont{Frixione}(1993)}]{Frixione:1993yp}
\bibinfo{author}{\bibfnamefont{S.}~\bibnamefont{Frixione}},
  \bibinfo{journal}{Nucl. Phys.} \textbf{\bibinfo{volume}{B410}},
  \bibinfo{pages}{280} (\bibinfo{year}{1993}).

\bibitem[{\citenamefont{Campbell et~al.}(2011)\citenamefont{Campbell, Ellis,
  and Williams}}]{Campbell:2011bn}
\bibinfo{author}{\bibfnamefont{J.~M.} \bibnamefont{Campbell}},
  \bibinfo{author}{\bibfnamefont{R.~K.} \bibnamefont{Ellis}}, \bibnamefont{and}
  \bibinfo{author}{\bibfnamefont{C.}~\bibnamefont{Williams}},
  \bibinfo{journal}{JHEP} \textbf{\bibinfo{volume}{07}}, \bibinfo{pages}{018}
  (\bibinfo{year}{2011}), \eprint{1105.0020}.

\bibitem[{\citenamefont{Dawson et~al.}(2013)\citenamefont{Dawson, Lewis, and
  Zeng}}]{Dawson:2013lya}
\bibinfo{author}{\bibfnamefont{S.}~\bibnamefont{Dawson}},
  \bibinfo{author}{\bibfnamefont{I.~M.} \bibnamefont{Lewis}}, \bibnamefont{and}
  \bibinfo{author}{\bibfnamefont{M.}~\bibnamefont{Zeng}},
  \bibinfo{journal}{Phys. Rev.} \textbf{\bibinfo{volume}{D88}},
  \bibinfo{pages}{054028} (\bibinfo{year}{2013}), \eprint{1307.3249}.

\bibitem[{\citenamefont{Aad et~al.}(2013)}]{ATLAS:2012mec}
\bibinfo{author}{\bibfnamefont{G.}~\bibnamefont{Aad}} \bibnamefont{et~al.}
  (\bibinfo{collaboration}{ATLAS}), \bibinfo{journal}{Phys. Rev.}
  \textbf{\bibinfo{volume}{D87}}, \bibinfo{pages}{112001}
  (\bibinfo{year}{2013}), \bibinfo{note}{[Erratum: Phys.
  Rev.D88,no.7,079906(2013)]}, \eprint{1210.2979}.

\bibitem[{ATL(2014)}]{ATLAS:2014xea}
\bibinfo{journal}{ATLAS-CONF-2014-033}  (\bibinfo{year}{2014}).

\bibitem[{CMS(2012{\natexlab{a}})}]{CMS:2012cva}
\bibinfo{journal}{CMS-PAS-SMP-12-005}  (\bibinfo{year}{2012}{\natexlab{a}}).

\bibitem[{CMS(2012{\natexlab{b}})}]{CMS:2012zva}
\bibinfo{journal}{CMS-PAS-SMP-12-013}  (\bibinfo{year}{2012}{\natexlab{b}}).

\bibitem[{\citenamefont{Curtin et~al.}(2013{\natexlab{a}})\citenamefont{Curtin,
  Jaiswal, Meade, and Tien}}]{Curtin:2013gta}
\bibinfo{author}{\bibfnamefont{D.}~\bibnamefont{Curtin}},
  \bibinfo{author}{\bibfnamefont{P.}~\bibnamefont{Jaiswal}},
  \bibinfo{author}{\bibfnamefont{P.}~\bibnamefont{Meade}}, \bibnamefont{and}
  \bibinfo{author}{\bibfnamefont{P.-J.} \bibnamefont{Tien}},
  \bibinfo{journal}{JHEP} \textbf{\bibinfo{volume}{08}}, \bibinfo{pages}{068}
  (\bibinfo{year}{2013}{\natexlab{a}}), \eprint{1304.7011}.

\bibitem[{\citenamefont{Curtin et~al.}(2013{\natexlab{b}})\citenamefont{Curtin,
  Jaiswal, and Meade}}]{Curtin:2012nn}
\bibinfo{author}{\bibfnamefont{D.}~\bibnamefont{Curtin}},
  \bibinfo{author}{\bibfnamefont{P.}~\bibnamefont{Jaiswal}}, \bibnamefont{and}
  \bibinfo{author}{\bibfnamefont{P.}~\bibnamefont{Meade}},
  \bibinfo{journal}{Phys. Rev.} \textbf{\bibinfo{volume}{D87}},
  \bibinfo{pages}{031701} (\bibinfo{year}{2013}{\natexlab{b}}),
  \eprint{1206.6888}.

\bibitem[{\citenamefont{Jaiswal et~al.}(2013)\citenamefont{Jaiswal, Kopp, and
  Okui}}]{Jaiswal:2013xra}
\bibinfo{author}{\bibfnamefont{P.}~\bibnamefont{Jaiswal}},
  \bibinfo{author}{\bibfnamefont{K.}~\bibnamefont{Kopp}}, \bibnamefont{and}
  \bibinfo{author}{\bibfnamefont{T.}~\bibnamefont{Okui}},
  \bibinfo{journal}{Phys. Rev.} \textbf{\bibinfo{volume}{D87}},
  \bibinfo{pages}{115017} (\bibinfo{year}{2013}), \eprint{1303.1181}.

\bibitem[{\citenamefont{Rolbiecki and Sakurai}(2013)}]{Rolbiecki:2013fia}
\bibinfo{author}{\bibfnamefont{K.}~\bibnamefont{Rolbiecki}} \bibnamefont{and}
  \bibinfo{author}{\bibfnamefont{K.}~\bibnamefont{Sakurai}},
  \bibinfo{journal}{JHEP} \textbf{\bibinfo{volume}{09}}, \bibinfo{pages}{004}
  (\bibinfo{year}{2013}), \eprint{1303.5696}.

\bibitem[{\citenamefont{Kim et~al.}(2014)\citenamefont{Kim, Rolbiecki, Sakurai,
  and Tattersall}}]{Kim:2014eva}
\bibinfo{author}{\bibfnamefont{J.~S.} \bibnamefont{Kim}},
  \bibinfo{author}{\bibfnamefont{K.}~\bibnamefont{Rolbiecki}},
  \bibinfo{author}{\bibfnamefont{K.}~\bibnamefont{Sakurai}}, \bibnamefont{and}
  \bibinfo{author}{\bibfnamefont{J.}~\bibnamefont{Tattersall}},
  \bibinfo{journal}{JHEP} \textbf{\bibinfo{volume}{12}}, \bibinfo{pages}{010}
  (\bibinfo{year}{2014}), \eprint{1406.0858}.

\bibitem[{\citenamefont{Curtin et~al.}(2014)\citenamefont{Curtin, Meade, and
  Tien}}]{Curtin:2014zua}
\bibinfo{author}{\bibfnamefont{D.}~\bibnamefont{Curtin}},
  \bibinfo{author}{\bibfnamefont{P.}~\bibnamefont{Meade}}, \bibnamefont{and}
  \bibinfo{author}{\bibfnamefont{P.-J.} \bibnamefont{Tien}},
  \bibinfo{journal}{Phys. Rev.} \textbf{\bibinfo{volume}{D90}},
  \bibinfo{pages}{115012} (\bibinfo{year}{2014}), \eprint{1406.0848}.

\bibitem[{\citenamefont{Khachatryan et~al.}(2016)}]{Khachatryan:2015sga}
\bibinfo{author}{\bibfnamefont{V.}~\bibnamefont{Khachatryan}}
  \bibnamefont{et~al.} (\bibinfo{collaboration}{CMS}), \bibinfo{journal}{Eur.
  Phys. J.} \textbf{\bibinfo{volume}{C76}}, \bibinfo{pages}{401}
  (\bibinfo{year}{2016}), \eprint{1507.03268}.

\bibitem[{\citenamefont{Aad et~al.}(2016)}]{Aad:2016wpd}
\bibinfo{author}{\bibfnamefont{G.}~\bibnamefont{Aad}} \bibnamefont{et~al.}
  (\bibinfo{collaboration}{ATLAS}) (\bibinfo{year}{2016}), \eprint{1603.01702}.

\bibitem[{ATL(2016)}]{ATLAS:2016rin}
\bibinfo{journal}{ATLAS-CONF-2016-090}  (\bibinfo{year}{2016}).

\bibitem[{CMS(2016)}]{CMS:2016vww}
\bibinfo{journal}{CMS-PAS-SMP-16-006}  (\bibinfo{year}{2016}).

\bibitem[{\citenamefont{Nason and Zanderighi}(2014)}]{Nason:2013ydw}
\bibinfo{author}{\bibfnamefont{P.}~\bibnamefont{Nason}} \bibnamefont{and}
  \bibinfo{author}{\bibfnamefont{G.}~\bibnamefont{Zanderighi}},
  \bibinfo{journal}{Eur. Phys. J.} \textbf{\bibinfo{volume}{C74}},
  \bibinfo{pages}{2702} (\bibinfo{year}{2014}), \eprint{1311.1365}.

\bibitem[{\citenamefont{Meade et~al.}(2014)\citenamefont{Meade, Ramani, and
  Zeng}}]{Meade:2014fca}
\bibinfo{author}{\bibfnamefont{P.}~\bibnamefont{Meade}},
  \bibinfo{author}{\bibfnamefont{H.}~\bibnamefont{Ramani}}, \bibnamefont{and}
  \bibinfo{author}{\bibfnamefont{M.}~\bibnamefont{Zeng}},
  \bibinfo{journal}{Phys. Rev.} \textbf{\bibinfo{volume}{D90}},
  \bibinfo{pages}{114006} (\bibinfo{year}{2014}), \eprint{1407.4481}.

\bibitem[{\citenamefont{Jaiswal and Okui}(2014)}]{Jaiswal:2014yba}
\bibinfo{author}{\bibfnamefont{P.}~\bibnamefont{Jaiswal}} \bibnamefont{and}
  \bibinfo{author}{\bibfnamefont{T.}~\bibnamefont{Okui}},
  \bibinfo{journal}{Phys. Rev.} \textbf{\bibinfo{volume}{D90}},
  \bibinfo{pages}{073009} (\bibinfo{year}{2014}), \eprint{1407.4537}.

\bibitem[{\citenamefont{Monni and Zanderighi}(2015)}]{Monni:2014zra}
\bibinfo{author}{\bibfnamefont{P.~F.} \bibnamefont{Monni}} \bibnamefont{and}
  \bibinfo{author}{\bibfnamefont{G.}~\bibnamefont{Zanderighi}},
  \bibinfo{journal}{JHEP} \textbf{\bibinfo{volume}{05}}, \bibinfo{pages}{013}
  (\bibinfo{year}{2015}), \eprint{1410.4745}.

\bibitem[{\citenamefont{Becher et~al.}(2015)\citenamefont{Becher, Frederix,
  Neubert, and Rothen}}]{Becher:2014aya}
\bibinfo{author}{\bibfnamefont{T.}~\bibnamefont{Becher}},
  \bibinfo{author}{\bibfnamefont{R.}~\bibnamefont{Frederix}},
  \bibinfo{author}{\bibfnamefont{M.}~\bibnamefont{Neubert}}, \bibnamefont{and}
  \bibinfo{author}{\bibfnamefont{L.}~\bibnamefont{Rothen}},
  \bibinfo{journal}{Eur. Phys. J.} \textbf{\bibinfo{volume}{C75}},
  \bibinfo{pages}{154} (\bibinfo{year}{2015}), \eprint{1412.8408}.

\bibitem[{\citenamefont{Jaiswal et~al.}(2016)\citenamefont{Jaiswal, Meade, and
  Ramani}}]{Jaiswal:2015vda}
\bibinfo{author}{\bibfnamefont{P.}~\bibnamefont{Jaiswal}},
  \bibinfo{author}{\bibfnamefont{P.}~\bibnamefont{Meade}}, \bibnamefont{and}
  \bibinfo{author}{\bibfnamefont{H.}~\bibnamefont{Ramani}},
  \bibinfo{journal}{Phys. Rev.} \textbf{\bibinfo{volume}{D93}},
  \bibinfo{pages}{093007} (\bibinfo{year}{2016}), \eprint{1509.07118}.

\bibitem[{\citenamefont{Ramani}(2015)}]{Ramani:2015mcd}
\bibinfo{author}{\bibfnamefont{H.}~\bibnamefont{Ramani}},
  \bibinfo{journal}{PoS} \textbf{\bibinfo{volume}{DIS2015}},
  \bibinfo{pages}{118} (\bibinfo{year}{2015}).

\bibitem[{\citenamefont{Gehrmann et~al.}(2015)\citenamefont{Gehrmann, von
  Manteuffel, and Tancredi}}]{Gehrmann:2015ora}
\bibinfo{author}{\bibfnamefont{T.}~\bibnamefont{Gehrmann}},
  \bibinfo{author}{\bibfnamefont{A.}~\bibnamefont{von Manteuffel}},
  \bibnamefont{and} \bibinfo{author}{\bibfnamefont{L.}~\bibnamefont{Tancredi}},
  \bibinfo{journal}{JHEP} \textbf{\bibinfo{volume}{09}}, \bibinfo{pages}{128}
  (\bibinfo{year}{2015}), \eprint{1503.04812}.

\bibitem[{\citenamefont{Caola et~al.}(2015)\citenamefont{Caola, Henn, Melnikov,
  Smirnov, and Smirnov}}]{Caola:2015ila}
\bibinfo{author}{\bibfnamefont{F.}~\bibnamefont{Caola}},
  \bibinfo{author}{\bibfnamefont{J.~M.} \bibnamefont{Henn}},
  \bibinfo{author}{\bibfnamefont{K.}~\bibnamefont{Melnikov}},
  \bibinfo{author}{\bibfnamefont{A.~V.} \bibnamefont{Smirnov}},
  \bibnamefont{and} \bibinfo{author}{\bibfnamefont{V.~A.}
  \bibnamefont{Smirnov}}, \bibinfo{journal}{JHEP}
  \textbf{\bibinfo{volume}{06}}, \bibinfo{pages}{129} (\bibinfo{year}{2015}),
  \eprint{1503.08759}.

\bibitem[{\citenamefont{Grazzini et~al.}(2015)\citenamefont{Grazzini, Kallweit,
  Rathlev, and Wiesemann}}]{Grazzini:2015wpa}
\bibinfo{author}{\bibfnamefont{M.}~\bibnamefont{Grazzini}},
  \bibinfo{author}{\bibfnamefont{S.}~\bibnamefont{Kallweit}},
  \bibinfo{author}{\bibfnamefont{D.}~\bibnamefont{Rathlev}}, \bibnamefont{and}
  \bibinfo{author}{\bibfnamefont{M.}~\bibnamefont{Wiesemann}},
  \bibinfo{journal}{JHEP} \textbf{\bibinfo{volume}{08}}, \bibinfo{pages}{154}
  (\bibinfo{year}{2015}), \eprint{1507.02565}.

\bibitem[{\citenamefont{Bauer et~al.}(2000)\citenamefont{Bauer, Fleming, and
  Luke}}]{Bauer:2000ew}
\bibinfo{author}{\bibfnamefont{C.~W.} \bibnamefont{Bauer}},
  \bibinfo{author}{\bibfnamefont{S.}~\bibnamefont{Fleming}}, \bibnamefont{and}
  \bibinfo{author}{\bibfnamefont{M.~E.} \bibnamefont{Luke}},
  \bibinfo{journal}{Phys. Rev.} \textbf{\bibinfo{volume}{D63}},
  \bibinfo{pages}{014006} (\bibinfo{year}{2000}), \eprint{hep-ph/0005275}.

\bibitem[{\citenamefont{Bauer et~al.}(2001)\citenamefont{Bauer, Fleming,
  Pirjol, and Stewart}}]{Bauer:2000yr}
\bibinfo{author}{\bibfnamefont{C.~W.} \bibnamefont{Bauer}},
  \bibinfo{author}{\bibfnamefont{S.}~\bibnamefont{Fleming}},
  \bibinfo{author}{\bibfnamefont{D.}~\bibnamefont{Pirjol}}, \bibnamefont{and}
  \bibinfo{author}{\bibfnamefont{I.~W.} \bibnamefont{Stewart}},
  \bibinfo{journal}{Phys. Rev.} \textbf{\bibinfo{volume}{D63}},
  \bibinfo{pages}{114020} (\bibinfo{year}{2001}), \eprint{hep-ph/0011336}.

\bibitem[{\citenamefont{Bauer and Stewart}(2001)}]{Bauer:2001ct}
\bibinfo{author}{\bibfnamefont{C.~W.} \bibnamefont{Bauer}} \bibnamefont{and}
  \bibinfo{author}{\bibfnamefont{I.~W.} \bibnamefont{Stewart}},
  \bibinfo{journal}{Phys. Lett.} \textbf{\bibinfo{volume}{B516}},
  \bibinfo{pages}{134} (\bibinfo{year}{2001}), \eprint{hep-ph/0107001}.

\bibitem[{\citenamefont{Bauer et~al.}(2002)\citenamefont{Bauer, Pirjol, and
  Stewart}}]{Bauer:2001yt}
\bibinfo{author}{\bibfnamefont{C.~W.} \bibnamefont{Bauer}},
  \bibinfo{author}{\bibfnamefont{D.}~\bibnamefont{Pirjol}}, \bibnamefont{and}
  \bibinfo{author}{\bibfnamefont{I.~W.} \bibnamefont{Stewart}},
  \bibinfo{journal}{Phys. Rev.} \textbf{\bibinfo{volume}{D65}},
  \bibinfo{pages}{054022} (\bibinfo{year}{2002}), \eprint{hep-ph/0109045}.

\bibitem[{\citenamefont{Beneke et~al.}(2002)\citenamefont{Beneke, Chapovsky,
  Diehl, and Feldmann}}]{Beneke:2002ph}
\bibinfo{author}{\bibfnamefont{M.}~\bibnamefont{Beneke}},
  \bibinfo{author}{\bibfnamefont{A.~P.} \bibnamefont{Chapovsky}},
  \bibinfo{author}{\bibfnamefont{M.}~\bibnamefont{Diehl}}, \bibnamefont{and}
  \bibinfo{author}{\bibfnamefont{T.}~\bibnamefont{Feldmann}},
  \bibinfo{journal}{Nucl. Phys.} \textbf{\bibinfo{volume}{B643}},
  \bibinfo{pages}{431} (\bibinfo{year}{2002}), \eprint{hep-ph/0206152}.

\bibitem[{\citenamefont{Beneke and Feldmann}(2003)}]{Beneke:2002ni}
\bibinfo{author}{\bibfnamefont{M.}~\bibnamefont{Beneke}} \bibnamefont{and}
  \bibinfo{author}{\bibfnamefont{T.}~\bibnamefont{Feldmann}},
  \bibinfo{journal}{Phys. Lett.} \textbf{\bibinfo{volume}{B553}},
  \bibinfo{pages}{267} (\bibinfo{year}{2003}), \eprint{hep-ph/0211358}.

\bibitem[{\citenamefont{Hill and Neubert}(2003)}]{Hill:2002vw}
\bibinfo{author}{\bibfnamefont{R.~J.} \bibnamefont{Hill}} \bibnamefont{and}
  \bibinfo{author}{\bibfnamefont{M.}~\bibnamefont{Neubert}},
  \bibinfo{journal}{Nucl. Phys.} \textbf{\bibinfo{volume}{B657}},
  \bibinfo{pages}{229} (\bibinfo{year}{2003}), \eprint{hep-ph/0211018}.

\bibitem[{\citenamefont{Binoth et~al.}(2006)\citenamefont{Binoth, Ciccolini,
  Kauer, and Kramer}}]{Binoth:2006mf}
\bibinfo{author}{\bibfnamefont{T.}~\bibnamefont{Binoth}},
  \bibinfo{author}{\bibfnamefont{M.}~\bibnamefont{Ciccolini}},
  \bibinfo{author}{\bibfnamefont{N.}~\bibnamefont{Kauer}}, \bibnamefont{and}
  \bibinfo{author}{\bibfnamefont{M.}~\bibnamefont{Kramer}},
  \bibinfo{journal}{JHEP} \textbf{\bibinfo{volume}{12}}, \bibinfo{pages}{046}
  (\bibinfo{year}{2006}), \eprint{hep-ph/0611170}.

\bibitem[{\citenamefont{Glover and van~der Bij}(1989)}]{Glover:1988fe}
\bibinfo{author}{\bibfnamefont{E.~W.~N.} \bibnamefont{Glover}}
  \bibnamefont{and} \bibinfo{author}{\bibfnamefont{J.~J.} \bibnamefont{van~der
  Bij}}, \bibinfo{journal}{Phys. Lett.} \textbf{\bibinfo{volume}{B219}},
  \bibinfo{pages}{488} (\bibinfo{year}{1989}).

\bibitem[{\citenamefont{Dicus et~al.}(1987)\citenamefont{Dicus, Kao, and
  Repko}}]{Dicus:1987dj}
\bibinfo{author}{\bibfnamefont{D.~A.} \bibnamefont{Dicus}},
  \bibinfo{author}{\bibfnamefont{C.}~\bibnamefont{Kao}}, \bibnamefont{and}
  \bibinfo{author}{\bibfnamefont{W.~W.} \bibnamefont{Repko}},
  \bibinfo{journal}{Phys. Rev.} \textbf{\bibinfo{volume}{D36}},
  \bibinfo{pages}{1570} (\bibinfo{year}{1987}).

\bibitem[{\citenamefont{Caola et~al.}(2016)\citenamefont{Caola, Melnikov,
  Rontsch, and Tancredi}}]{Caola:2015rqy}
\bibinfo{author}{\bibfnamefont{F.}~\bibnamefont{Caola}},
  \bibinfo{author}{\bibfnamefont{K.}~\bibnamefont{Melnikov}},
  \bibinfo{author}{\bibfnamefont{R.}~\bibnamefont{Rontsch}}, \bibnamefont{and}
  \bibinfo{author}{\bibfnamefont{L.}~\bibnamefont{Tancredi}},
  \bibinfo{journal}{Phys. Lett.} \textbf{\bibinfo{volume}{B754}},
  \bibinfo{pages}{275} (\bibinfo{year}{2016}), \eprint{1511.08617}.

\bibitem[{\citenamefont{Gleisberg et~al.}(2009)\citenamefont{Gleisberg, Hoeche,
  Krauss, Schonherr, Schumann, Siegert, and Winter}}]{Gleisberg:2008ta}
\bibinfo{author}{\bibfnamefont{T.}~\bibnamefont{Gleisberg}},
  \bibinfo{author}{\bibfnamefont{S.}~\bibnamefont{Hoeche}},
  \bibinfo{author}{\bibfnamefont{F.}~\bibnamefont{Krauss}},
  \bibinfo{author}{\bibfnamefont{M.}~\bibnamefont{Schonherr}},
  \bibinfo{author}{\bibfnamefont{S.}~\bibnamefont{Schumann}},
  \bibinfo{author}{\bibfnamefont{F.}~\bibnamefont{Siegert}}, \bibnamefont{and}
  \bibinfo{author}{\bibfnamefont{J.}~\bibnamefont{Winter}},
  \bibinfo{journal}{JHEP} \textbf{\bibinfo{volume}{02}}, \bibinfo{pages}{007}
  (\bibinfo{year}{2009}), \eprint{0811.4622}.

\bibitem[{\citenamefont{Catani and Grazzini}(2007)}]{Catani:2007vq}
\bibinfo{author}{\bibfnamefont{S.}~\bibnamefont{Catani}} \bibnamefont{and}
  \bibinfo{author}{\bibfnamefont{M.}~\bibnamefont{Grazzini}},
  \bibinfo{journal}{Phys. Rev. Lett.} \textbf{\bibinfo{volume}{98}},
  \bibinfo{pages}{222002} (\bibinfo{year}{2007}), \eprint{hep-ph/0703012}.

\bibitem[{\citenamefont{Krauss et~al.}(2002)\citenamefont{Krauss, Kuhn, and
  Soff}}]{Krauss:2001iv}
\bibinfo{author}{\bibfnamefont{F.}~\bibnamefont{Krauss}},
  \bibinfo{author}{\bibfnamefont{R.}~\bibnamefont{Kuhn}}, \bibnamefont{and}
  \bibinfo{author}{\bibfnamefont{G.}~\bibnamefont{Soff}},
  \bibinfo{journal}{JHEP} \textbf{\bibinfo{volume}{02}}, \bibinfo{pages}{044}
  (\bibinfo{year}{2002}), \eprint{hep-ph/0109036}.

\bibitem[{\citenamefont{Gleisberg and Hoeche}(2008)}]{Gleisberg:2008fv}
\bibinfo{author}{\bibfnamefont{T.}~\bibnamefont{Gleisberg}} \bibnamefont{and}
  \bibinfo{author}{\bibfnamefont{S.}~\bibnamefont{Hoeche}},
  \bibinfo{journal}{JHEP} \textbf{\bibinfo{volume}{12}}, \bibinfo{pages}{039}
  (\bibinfo{year}{2008}), \eprint{0808.3674}.

\bibitem[{\citenamefont{Becher and Neubert}(2011)}]{Becher:2010tm}
\bibinfo{author}{\bibfnamefont{T.}~\bibnamefont{Becher}} \bibnamefont{and}
  \bibinfo{author}{\bibfnamefont{M.}~\bibnamefont{Neubert}},
  \bibinfo{journal}{Eur. Phys. J.} \textbf{\bibinfo{volume}{C71}},
  \bibinfo{pages}{1665} (\bibinfo{year}{2011}), \eprint{1007.4005}.

\bibitem[{\citenamefont{Gehrmann
  et~al.}(2014{\natexlab{b}})\citenamefont{Gehrmann, Luebbert, and
  Yang}}]{Gehrmann:2014yya}
\bibinfo{author}{\bibfnamefont{T.}~\bibnamefont{Gehrmann}},
  \bibinfo{author}{\bibfnamefont{T.}~\bibnamefont{Luebbert}}, \bibnamefont{and}
  \bibinfo{author}{\bibfnamefont{L.~L.} \bibnamefont{Yang}},
  \bibinfo{journal}{JHEP} \textbf{\bibinfo{volume}{06}}, \bibinfo{pages}{155}
  (\bibinfo{year}{2014}{\natexlab{b}}), \eprint{1403.6451}.

\bibitem[{\citenamefont{Dittmaier et~al.}(2008)\citenamefont{Dittmaier,
  Kallweit, and Uwer}}]{Dittmaier:2007th}
\bibinfo{author}{\bibfnamefont{S.}~\bibnamefont{Dittmaier}},
  \bibinfo{author}{\bibfnamefont{S.}~\bibnamefont{Kallweit}}, \bibnamefont{and}
  \bibinfo{author}{\bibfnamefont{P.}~\bibnamefont{Uwer}},
  \bibinfo{journal}{Phys. Rev. Lett.} \textbf{\bibinfo{volume}{100}},
  \bibinfo{pages}{062003} (\bibinfo{year}{2008}), \eprint{0710.1577}.

\bibitem[{\citenamefont{Melia et~al.}(2012)\citenamefont{Melia, Melnikov,
  Rontsch, Schulze, and Zanderighi}}]{Melia:2012zg}
\bibinfo{author}{\bibfnamefont{T.}~\bibnamefont{Melia}},
  \bibinfo{author}{\bibfnamefont{K.}~\bibnamefont{Melnikov}},
  \bibinfo{author}{\bibfnamefont{R.}~\bibnamefont{Rontsch}},
  \bibinfo{author}{\bibfnamefont{M.}~\bibnamefont{Schulze}}, \bibnamefont{and}
  \bibinfo{author}{\bibfnamefont{G.}~\bibnamefont{Zanderighi}},
  \bibinfo{journal}{JHEP} \textbf{\bibinfo{volume}{08}}, \bibinfo{pages}{115}
  (\bibinfo{year}{2012}), \eprint{1205.6987}.

\bibitem[{\citenamefont{Campbell et~al.}(2015)\citenamefont{Campbell, Miller,
  and Robens}}]{Campbell:2015hya}
\bibinfo{author}{\bibfnamefont{J.~M.} \bibnamefont{Campbell}},
  \bibinfo{author}{\bibfnamefont{D.~J.} \bibnamefont{Miller}},
  \bibnamefont{and} \bibinfo{author}{\bibfnamefont{T.}~\bibnamefont{Robens}},
  \bibinfo{journal}{Phys. Rev.} \textbf{\bibinfo{volume}{D92}},
  \bibinfo{pages}{014033} (\bibinfo{year}{2015}), \eprint{1506.04801}.

\bibitem[{\citenamefont{Catani and Seymour}(1997)}]{Catani:1996vz}
\bibinfo{author}{\bibfnamefont{S.}~\bibnamefont{Catani}} \bibnamefont{and}
  \bibinfo{author}{\bibfnamefont{M.~H.} \bibnamefont{Seymour}},
  \bibinfo{journal}{Nucl. Phys.} \textbf{\bibinfo{volume}{B485}},
  \bibinfo{pages}{291} (\bibinfo{year}{1997}), \bibinfo{note}{[Erratum: Nucl.
  Phys.B510,503(1998)]}, \eprint{hep-ph/9605323}.

\bibitem[{\citenamefont{Cascioli et~al.}(2012)\citenamefont{Cascioli,
  Maierhofer, and Pozzorini}}]{Cascioli:2011va}
\bibinfo{author}{\bibfnamefont{F.}~\bibnamefont{Cascioli}},
  \bibinfo{author}{\bibfnamefont{P.}~\bibnamefont{Maierhofer}},
  \bibnamefont{and}
  \bibinfo{author}{\bibfnamefont{S.}~\bibnamefont{Pozzorini}},
  \bibinfo{journal}{Phys. Rev. Lett.} \textbf{\bibinfo{volume}{108}},
  \bibinfo{pages}{111601} (\bibinfo{year}{2012}), \eprint{1111.5206}.

\bibitem[{\citenamefont{Catani et~al.}(2014)\citenamefont{Catani, Cieri,
  de~Florian, Ferrera, and Grazzini}}]{Catani:2013tia}
\bibinfo{author}{\bibfnamefont{S.}~\bibnamefont{Catani}},
  \bibinfo{author}{\bibfnamefont{L.}~\bibnamefont{Cieri}},
  \bibinfo{author}{\bibfnamefont{D.}~\bibnamefont{de~Florian}},
  \bibinfo{author}{\bibfnamefont{G.}~\bibnamefont{Ferrera}}, \bibnamefont{and}
  \bibinfo{author}{\bibfnamefont{M.}~\bibnamefont{Grazzini}},
  \bibinfo{journal}{Nucl. Phys.} \textbf{\bibinfo{volume}{B881}},
  \bibinfo{pages}{414} (\bibinfo{year}{2014}), \eprint{1311.1654}.

\bibitem[{\citenamefont{Anastasiou et~al.}(2003)\citenamefont{Anastasiou,
  Dixon, Melnikov, and Petriello}}]{Anastasiou:2003yy}
\bibinfo{author}{\bibfnamefont{C.}~\bibnamefont{Anastasiou}},
  \bibinfo{author}{\bibfnamefont{L.~J.} \bibnamefont{Dixon}},
  \bibinfo{author}{\bibfnamefont{K.}~\bibnamefont{Melnikov}}, \bibnamefont{and}
  \bibinfo{author}{\bibfnamefont{F.}~\bibnamefont{Petriello}},
  \bibinfo{journal}{Phys. Rev. Lett.} \textbf{\bibinfo{volume}{91}},
  \bibinfo{pages}{182002} (\bibinfo{year}{2003}), \eprint{hep-ph/0306192}.

\bibitem[{\citenamefont{Anastasiou et~al.}(2002)\citenamefont{Anastasiou,
  Glover, and Tejeda-Yeomans}}]{Anastasiou:2002zn}
\bibinfo{author}{\bibfnamefont{C.}~\bibnamefont{Anastasiou}},
  \bibinfo{author}{\bibfnamefont{E.~W.~N.} \bibnamefont{Glover}},
  \bibnamefont{and} \bibinfo{author}{\bibfnamefont{M.~E.}
  \bibnamefont{Tejeda-Yeomans}}, \bibinfo{journal}{Nucl. Phys.}
  \textbf{\bibinfo{volume}{B629}}, \bibinfo{pages}{255} (\bibinfo{year}{2002}),
  \eprint{hep-ph/0201274}.

\bibitem[{\citenamefont{Martin et~al.}(2010)\citenamefont{Martin, Stirling,
  Thorne, and Watt}}]{Martin:2010db}
\bibinfo{author}{\bibfnamefont{A.~D.} \bibnamefont{Martin}},
  \bibinfo{author}{\bibfnamefont{W.~J.} \bibnamefont{Stirling}},
  \bibinfo{author}{\bibfnamefont{R.~S.} \bibnamefont{Thorne}},
  \bibnamefont{and} \bibinfo{author}{\bibfnamefont{G.}~\bibnamefont{Watt}},
  \bibinfo{journal}{Eur. Phys. J.} \textbf{\bibinfo{volume}{C70}},
  \bibinfo{pages}{51} (\bibinfo{year}{2010}), \eprint{1007.2624}.

\bibitem[{\citenamefont{Banfi et~al.}(2012{\natexlab{a}})\citenamefont{Banfi,
  Salam, and Zanderighi}}]{Banfi:2012yh}
\bibinfo{author}{\bibfnamefont{A.}~\bibnamefont{Banfi}},
  \bibinfo{author}{\bibfnamefont{G.~P.} \bibnamefont{Salam}}, \bibnamefont{and}
  \bibinfo{author}{\bibfnamefont{G.}~\bibnamefont{Zanderighi}},
  \bibinfo{journal}{JHEP} \textbf{\bibinfo{volume}{06}}, \bibinfo{pages}{159}
  (\bibinfo{year}{2012}{\natexlab{a}}), \eprint{1203.5773}.

\bibitem[{\citenamefont{Banfi et~al.}(2012{\natexlab{b}})\citenamefont{Banfi,
  Monni, Salam, and Zanderighi}}]{Banfi:2012jm}
\bibinfo{author}{\bibfnamefont{A.}~\bibnamefont{Banfi}},
  \bibinfo{author}{\bibfnamefont{P.~F.} \bibnamefont{Monni}},
  \bibinfo{author}{\bibfnamefont{G.~P.} \bibnamefont{Salam}}, \bibnamefont{and}
  \bibinfo{author}{\bibfnamefont{G.}~\bibnamefont{Zanderighi}},
  \bibinfo{journal}{Phys. Rev. Lett.} \textbf{\bibinfo{volume}{109}},
  \bibinfo{pages}{202001} (\bibinfo{year}{2012}{\natexlab{b}}),
  \eprint{1206.4998}.

\bibitem[{\citenamefont{Berger et~al.}(2011)\citenamefont{Berger, Marcantonini,
  Stewart, Tackmann, and Waalewijn}}]{Berger:2010xi}
\bibinfo{author}{\bibfnamefont{C.~F.} \bibnamefont{Berger}},
  \bibinfo{author}{\bibfnamefont{C.}~\bibnamefont{Marcantonini}},
  \bibinfo{author}{\bibfnamefont{I.~W.} \bibnamefont{Stewart}},
  \bibinfo{author}{\bibfnamefont{F.~J.} \bibnamefont{Tackmann}},
  \bibnamefont{and} \bibinfo{author}{\bibfnamefont{W.~J.}
  \bibnamefont{Waalewijn}}, \bibinfo{journal}{JHEP}
  \textbf{\bibinfo{volume}{04}}, \bibinfo{pages}{092} (\bibinfo{year}{2011}),
  \eprint{1012.4480}.

\bibitem[{\citenamefont{Tackmann et~al.}(2012)\citenamefont{Tackmann, Walsh,
  and Zuberi}}]{Tackmann:2012bt}
\bibinfo{author}{\bibfnamefont{F.~J.} \bibnamefont{Tackmann}},
  \bibinfo{author}{\bibfnamefont{J.~R.} \bibnamefont{Walsh}}, \bibnamefont{and}
  \bibinfo{author}{\bibfnamefont{S.}~\bibnamefont{Zuberi}},
  \bibinfo{journal}{Phys. Rev.} \textbf{\bibinfo{volume}{D86}},
  \bibinfo{pages}{053011} (\bibinfo{year}{2012}), \eprint{1206.4312}.

\bibitem[{\citenamefont{Stewart et~al.}(2014)\citenamefont{Stewart, Tackmann,
  Walsh, and Zuberi}}]{Stewart:2013faa}
\bibinfo{author}{\bibfnamefont{I.~W.} \bibnamefont{Stewart}},
  \bibinfo{author}{\bibfnamefont{F.~J.} \bibnamefont{Tackmann}},
  \bibinfo{author}{\bibfnamefont{J.~R.} \bibnamefont{Walsh}}, \bibnamefont{and}
  \bibinfo{author}{\bibfnamefont{S.}~\bibnamefont{Zuberi}},
  \bibinfo{journal}{Phys. Rev.} \textbf{\bibinfo{volume}{D89}},
  \bibinfo{pages}{054001} (\bibinfo{year}{2014}), \eprint{1307.1808}.

\bibitem[{\citenamefont{Becher and Neubert}(2012)}]{Becher:2012qa}
\bibinfo{author}{\bibfnamefont{T.}~\bibnamefont{Becher}} \bibnamefont{and}
  \bibinfo{author}{\bibfnamefont{M.}~\bibnamefont{Neubert}},
  \bibinfo{journal}{JHEP} \textbf{\bibinfo{volume}{07}}, \bibinfo{pages}{108}
  (\bibinfo{year}{2012}), \eprint{1205.3806}.

\bibitem[{\citenamefont{Becher et~al.}(2013)\citenamefont{Becher, Neubert, and
  Rothen}}]{Becher:2013xia}
\bibinfo{author}{\bibfnamefont{T.}~\bibnamefont{Becher}},
  \bibinfo{author}{\bibfnamefont{M.}~\bibnamefont{Neubert}}, \bibnamefont{and}
  \bibinfo{author}{\bibfnamefont{L.}~\bibnamefont{Rothen}},
  \bibinfo{journal}{JHEP} \textbf{\bibinfo{volume}{10}}, \bibinfo{pages}{125}
  (\bibinfo{year}{2013}), \eprint{1307.0025}.

\bibitem[{\citenamefont{Jaiswal and Okui}(2015)}]{Jaiswal:2015nka}
\bibinfo{author}{\bibfnamefont{P.}~\bibnamefont{Jaiswal}} \bibnamefont{and}
  \bibinfo{author}{\bibfnamefont{T.}~\bibnamefont{Okui}},
  \bibinfo{journal}{Phys. Rev.} \textbf{\bibinfo{volume}{D92}},
  \bibinfo{pages}{074035} (\bibinfo{year}{2015}), \eprint{1506.07529}.

\bibitem[{\citenamefont{Gaunt}(2014)}]{Gaunt:2014ska}
\bibinfo{author}{\bibfnamefont{J.~R.} \bibnamefont{Gaunt}},
  \bibinfo{journal}{JHEP} \textbf{\bibinfo{volume}{07}}, \bibinfo{pages}{110}
  (\bibinfo{year}{2014}), \eprint{1405.2080}.

\bibitem[{\citenamefont{Zeng}(2015)}]{Zeng:2015iba}
\bibinfo{author}{\bibfnamefont{M.}~\bibnamefont{Zeng}}, \bibinfo{journal}{JHEP}
  \textbf{\bibinfo{volume}{10}}, \bibinfo{pages}{189} (\bibinfo{year}{2015}),
  \eprint{1507.01652}.

\bibitem[{\citenamefont{Rothstein and Stewart}(2016)}]{Rothstein:2016bsq}
\bibinfo{author}{\bibfnamefont{I.~Z.} \bibnamefont{Rothstein}}
  \bibnamefont{and} \bibinfo{author}{\bibfnamefont{I.~W.}
  \bibnamefont{Stewart}} (\bibinfo{year}{2016}), \eprint{1601.04695}.

\bibitem[{\citenamefont{Becher et~al.}(2008)\citenamefont{Becher, Neubert, and
  Xu}}]{Becher:2007ty}
\bibinfo{author}{\bibfnamefont{T.}~\bibnamefont{Becher}},
  \bibinfo{author}{\bibfnamefont{M.}~\bibnamefont{Neubert}}, \bibnamefont{and}
  \bibinfo{author}{\bibfnamefont{G.}~\bibnamefont{Xu}}, \bibinfo{journal}{JHEP}
  \textbf{\bibinfo{volume}{07}}, \bibinfo{pages}{030} (\bibinfo{year}{2008}),
  \eprint{0710.0680}.

\bibitem[{\citenamefont{Li and Liu}(2014)}]{Li:2014ria}
\bibinfo{author}{\bibfnamefont{Y.}~\bibnamefont{Li}} \bibnamefont{and}
  \bibinfo{author}{\bibfnamefont{X.}~\bibnamefont{Liu}},
  \bibinfo{journal}{JHEP} \textbf{\bibinfo{volume}{06}}, \bibinfo{pages}{028}
  (\bibinfo{year}{2014}), \eprint{1401.2149}.

\bibitem[{\citenamefont{Dasgupta et~al.}(2015)\citenamefont{Dasgupta, Dreyer,
  Salam, and Soyez}}]{Dasgupta:2014yra}
\bibinfo{author}{\bibfnamefont{M.}~\bibnamefont{Dasgupta}},
  \bibinfo{author}{\bibfnamefont{F.}~\bibnamefont{Dreyer}},
  \bibinfo{author}{\bibfnamefont{G.~P.} \bibnamefont{Salam}}, \bibnamefont{and}
  \bibinfo{author}{\bibfnamefont{G.}~\bibnamefont{Soyez}},
  \bibinfo{journal}{JHEP} \textbf{\bibinfo{volume}{04}}, \bibinfo{pages}{039}
  (\bibinfo{year}{2015}), \eprint{1411.5182}.

\bibitem[{\citenamefont{Banfi et~al.}(2016)\citenamefont{Banfi, Caola, Dreyer,
  Monni, Salam, Zanderighi, and Dulat}}]{Banfi:2015pju}
\bibinfo{author}{\bibfnamefont{A.}~\bibnamefont{Banfi}},
  \bibinfo{author}{\bibfnamefont{F.}~\bibnamefont{Caola}},
  \bibinfo{author}{\bibfnamefont{F.~A.} \bibnamefont{Dreyer}},
  \bibinfo{author}{\bibfnamefont{P.~F.} \bibnamefont{Monni}},
  \bibinfo{author}{\bibfnamefont{G.~P.} \bibnamefont{Salam}},
  \bibinfo{author}{\bibfnamefont{G.}~\bibnamefont{Zanderighi}},
  \bibnamefont{and} \bibinfo{author}{\bibfnamefont{F.}~\bibnamefont{Dulat}},
  \bibinfo{journal}{JHEP} \textbf{\bibinfo{volume}{04}}, \bibinfo{pages}{049}
  (\bibinfo{year}{2016}), \eprint{1511.02886}.

\bibitem[{\citenamefont{Chiu et~al.}(2012{\natexlab{a}})\citenamefont{Chiu,
  Jain, Neill, and Rothstein}}]{Chiu:2011qc}
\bibinfo{author}{\bibfnamefont{J.-y.} \bibnamefont{Chiu}},
  \bibinfo{author}{\bibfnamefont{A.}~\bibnamefont{Jain}},
  \bibinfo{author}{\bibfnamefont{D.}~\bibnamefont{Neill}}, \bibnamefont{and}
  \bibinfo{author}{\bibfnamefont{I.~Z.} \bibnamefont{Rothstein}},
  \bibinfo{journal}{Phys. Rev. Lett.} \textbf{\bibinfo{volume}{108}},
  \bibinfo{pages}{151601} (\bibinfo{year}{2012}{\natexlab{a}}),
  \eprint{1104.0881}.

\bibitem[{\citenamefont{Chiu et~al.}(2012{\natexlab{b}})\citenamefont{Chiu,
  Jain, Neill, and Rothstein}}]{Chiu:2012ir}
\bibinfo{author}{\bibfnamefont{J.-Y.} \bibnamefont{Chiu}},
  \bibinfo{author}{\bibfnamefont{A.}~\bibnamefont{Jain}},
  \bibinfo{author}{\bibfnamefont{D.}~\bibnamefont{Neill}}, \bibnamefont{and}
  \bibinfo{author}{\bibfnamefont{I.~Z.} \bibnamefont{Rothstein}},
  \bibinfo{journal}{JHEP} \textbf{\bibinfo{volume}{05}}, \bibinfo{pages}{084}
  (\bibinfo{year}{2012}{\natexlab{b}}), \eprint{1202.0814}.

\bibitem[{\citenamefont{H{\"o}che et~al.}(2015)\citenamefont{H{\"o}che, Li, and
  Prestel}}]{Hoeche:2014aia}
\bibinfo{author}{\bibfnamefont{S.}~\bibnamefont{H{\"o}che}},
  \bibinfo{author}{\bibfnamefont{Y.}~\bibnamefont{Li}}, \bibnamefont{and}
  \bibinfo{author}{\bibfnamefont{S.}~\bibnamefont{Prestel}},
  \bibinfo{journal}{Phys. Rev.} \textbf{\bibinfo{volume}{D91}},
  \bibinfo{pages}{074015} (\bibinfo{year}{2015}), \eprint{1405.3607}.

\bibitem[{\citenamefont{H{\"o}che et~al.}(2014)\citenamefont{H{\"o}che, Li, and
  Prestel}}]{Hoche:2014dla}
\bibinfo{author}{\bibfnamefont{S.}~\bibnamefont{H{\"o}che}},
  \bibinfo{author}{\bibfnamefont{Y.}~\bibnamefont{Li}}, \bibnamefont{and}
  \bibinfo{author}{\bibfnamefont{S.}~\bibnamefont{Prestel}},
  \bibinfo{journal}{Phys. Rev.} \textbf{\bibinfo{volume}{D90}},
  \bibinfo{pages}{054011} (\bibinfo{year}{2014}), \eprint{1407.3773}.

\end{thebibliography}

\end{document}